\documentclass[12pt]{article} 
\usepackage{amsmath}
\usepackage{amssymb}
\usepackage{epsfig}

\newcommand{\be}{\begin{equation}}
\newcommand{\ee}{\end{equation}}
\newcommand{\ba}{\begin{eqnarray}}
\newcommand{\ea}{\end{eqnarray}}

\begin{document} 
\setlength{\parskip}{0.45cm} 
\setlength{\baselineskip}{0.75cm} 

%
%
%
\begin{titlepage} 
\setlength{\parskip}{0.25cm} 
\setlength{\baselineskip}{0.25cm}
\begin{flushright} 
DO-TH 2005/04 \\ 
\vspace{0.2cm} 
astro--ph/0503442\\ 
\vspace{0.2cm} 
March 2005 
\end{flushright} 
\vspace{1.0cm} 
\begin{center} 
\LARGE 
{\bf Alternative Multipole Vectors for the CMB Temperature Fluctuations} 
\vspace{1.5cm} 
 
\large 
M. Gl\"uck and C.\ Pisano\\
\vspace{1.0cm}

\normalsize 
{\it Universit\"{a}t Dortmund, Institut f\"{u}r Physik,}\\ 
{\it D-44221 Dortmund, Germany} \\ 
\vspace{0.5cm}

\vspace{1.5cm} 
\end{center} 
 
\begin{abstract} 
\noindent 
We introduce new, low-$\ell$, multipole vectors for the CMB temperature
fluctuations. A very strong alignment of the quadrupole and octopole
vectors is observed as well as a remarkably low declination of these
vectors with respect to the galactic plane.
\end{abstract} 
\end{titlepage} 
 

The temperature fluctuations, $\Delta T(\Omega)$, of the CMB radiation 
are traditionally decomposed according to
\be
\Delta T (\Omega) = \sum_{{\ell} m} a_{{\ell} m} Y_{{\ell} m}(\Omega)
\ee
where $a_{\ell -m} = (-1)^ma_{\ell m}^*$ reflects the reality of 
$\Delta T (\Omega )$. For each $\ell$ there are  thus $2\ell +1$ independent
real parameters which can be utilized in general to construct $\ell$ 
uncorrelated unit vectors and one common overall amplitude.

In \cite{Copi:2003kt} a method to construct such multipole vectors was 
introduced and the properties of these vectors were studied in 
\cite{Copi:2003kt,Schwarz:2004gk}. In particular, it turned out that the 
$\ell = 2$ and 3 multipole vectors possess quite unusual correlations which are
unexpected for a statistically isotropic sky with Gaussian random 
$a_{\ell m}$, $\langle a_{\ell m} a_{\ell m}^*\rangle = C_{\ell}\delta_{\ell
\ell '}\delta_{m m'}$. The multipole vectors presented in \cite{Copi:2003kt}
were obtained by the construction of symmetric traceless rank-$\ell$
tensor representation of the spherical harmonics $Y_{\ell m}$ and it remains 
to be seen 
to what extent the above mentioned correlations depend on these particularly 
chosen multipole vectors.
 
To study this question we introduce an alternative set of multipole vectors
whose directions point towards the maxima or minima of 
$\Delta T_{\ell}(\Omega) \equiv \sum_{m}a_{\ell m} Y_{\ell m}(\Omega)$.
Utilizing the $a_{\ell m}$ in Table II of \cite{deOliveira-Costa:2003pu},
obtained from a foreground cleaned \cite{Tegmark:2003ve} CMB map one
obtains in galactic coordinates $(l, b)$ the following unit vectors:
\ba
\hat{\mbox{\boldmath $n$}}_1  & = & (157.03^{\circ}, 6.29^{\circ}), 
\nonumber \\
\hat{\mbox{\boldmath $n$}}_2  & = & (63.33^{\circ},  30.41^{\circ}),
\ea
in the quadrupole sector and
\ba
\hat{\mbox{\boldmath $n$}}_a &=&(145.97^{\circ}, 10.59^{\circ}),
\nonumber \\ 
\hat{\mbox{\boldmath $n$}}_b &=& (14.59^{\circ}, 29.75^{\circ}), 
\nonumber \\
\hat{\mbox{\boldmath $n$}}_c &=& (82.09^{\circ}, 17.93^{\circ}),     
\ea
in the octopole sector. For convenience these vectors were chosen to point
towards the north galactic hemisphere. We note the rather unusual alignments
between some of the $\ell = 2$ and $\ell = 3$ vectors, i.e.:
\be
\hat{\mbox{\boldmath $n$}}_1 \cdot\hat{\mbox{\boldmath $n$}}_a  = 0.9787, ~~~~~~~~~~\hat{\mbox{\boldmath $n$}}_2\cdot\hat{\mbox{\boldmath $n$}}_c = 0.9328, 
\label{eq:scalar_a}
\ee
which in a statistically isotropic sky with random $a_{\ell m}$ is expected 
to happen by chance only about once in $(1-\hat{\mbox{\boldmath $n$}}_1\cdot\hat{\mbox{\boldmath $n$}}_a)^{-1}
(1-\hat{\mbox{\boldmath $n$}}_2\cdot\hat{\mbox{\boldmath $n$}}_c)^{-1} 
\simeq 700 $, (cf. footnote 4 in 
\cite{deOliveira-Costa:2003pu}). Similar alignments also exist for
the multipole vectors  $\hat{\mbox{\boldmath $v$}}^{(\ell , i)}$ of 
\cite{Schwarz:2004gk}:
\be
\hat{\mbox{\boldmath $v$}}^{(2, 1)} \cdot 
\hat{\mbox{\boldmath $v$}}^{(3,1)}=  0.9730, ~~~~~~~~~~~
\hat{\mbox{\boldmath $v$}}^{(2, 2)} \cdot 
\hat{\mbox{\boldmath $v$}}^{(3, 2)}= 0.8636, 
\label{eq:scalar}
\ee
which is expected to happen by chance only about once in 270, quite lower
than for the previous alignments. The coplanarity of our unit vectors is 
characterized by 
\be
D_{a b} = 0.9715, ~~~~ D_{ac} = 0.9693,~~~~D_{b c} = 0.9007,
\label{eq:copl} 
\ee
where, following \cite{Copi:2003kt,Schwarz:2004gk}, 
$D_{ab} \equiv |(\hat{\mbox{\boldmath $n$}}_1
\times\hat{\mbox{\boldmath $n$}}_2)
\cdot (\hat{\mbox{\boldmath $n$}}_a\times\hat{\mbox{\boldmath $n$}}_b) |/
|\hat{\mbox{\boldmath $n$}}_1 \times\hat{\mbox{\boldmath $n$}}_2 | 
|\hat{\mbox{\boldmath $n$}}_a\times\hat{\mbox{\boldmath $n$}}_b|$, with 
analogous definitions for $D_{a c}$ and $D_{b c}$.   
The corresponding products for the vectors 
$\hat{\mbox{\boldmath $v$}}^{(\ell, i)}$ are 
\cite{Schwarz:2004gk}:
\be
D_1 = 0.9531,~~~~D_2 = 0.8719,~~~~D_3 = 0.8377,
\ee
i.e. a somewhat reduced coplanarity between the $\ell = 2$ and $\ell = 3$
vectors as compared to (\ref{eq:copl}). This reduction is directly 
related to the corresponding reduced alignments in (\ref{eq:scalar})
as compared to (\ref{eq:scalar_a}).

Finally we note the lower declinations $b(\hat{\mbox{\boldmath $n$}})$ as
compared to $b(\hat{\mbox{\boldmath $v$}})$
implying a closer vicinity of our multipole vectors to the galactic plane.
This close vicinity, characterized by  $\cos b(\hat{\mbox{\boldmath $n$}}_1)= 
0.9940$, and the 
apparently unrelated but unusual alignments of our $\ell = 2$ and $\ell = 3$
unit vectors, as given in (\ref{eq:scalar_a}), suggest a common origin for 
these -- seemingly unrelated -- remarkable correlations.

To summarize, we have introduced a new set of $\ell = 2$, $3$ multipole vectors
pointing towards the maxima or minima of the corresponding CMB temperature 
fluctuations. These vectors are characterized by an unusual high 
alignment between
some quadrupole and octopole unit vectors as well as by a simultaneous 
very low declination of these vectors with respect to the galactic
plane.\\

This work has been supported in part by the `Bundesministerium f\"ur
Bildung und Forschung', Berlin/Bonn.



\begin{thebibliography}{99}

\bibitem{Copi:2003kt}
  C.~J.~Copi, D.~Huterer and G.~D.~Starkman,
  Phys.\ Rev.\ D {\bf 70}, 043515 (2004).

\bibitem{Schwarz:2004gk}
  D.~J.~Schwarz, G.~D.~Starkman, D.~Huterer and C.~J.~Copi,
  Phys.\ Rev.\ Lett.\  {\bf 93}, 221301 (2004).

\bibitem{deOliveira-Costa:2003pu}
  A.~de Oliveira-Costa, M.~Tegmark, M.~Zaldarriaga and A.~Hamilton,
  Phys.\ Rev.\ D {\bf 69}, 063516 (2004).


\bibitem{Tegmark:2003ve}
  M.~Tegmark, A.~de Oliveira-Costa and A.~Hamilton,
  Phys.\ Rev.\ D {\bf 68}, 123523 (2003).







\end{thebibliography}
\end{document}